\documentclass[reprint,amsmath,amssymb,aps]{revtex4-2}
\usepackage[dvipdfmx]{hyperref,graphicx}
\usepackage{dcolumn}
\usepackage{bm}
\usepackage{physics}
\usepackage{appendix}
\usepackage{comment}
\usepackage{mathtools}
\usepackage{here}
\usepackage{tikz}
\usepackage{color}
\usepackage{ulem}

\hypersetup{
	colorlinks=false, 
	bookmarks=true, 
	bookmarksnumbered=true,
	pdfborder={0 0 0},
	bookmarkstype=toc
}

\newcommand{\wt}[1]{\widetilde{#1}}

\begin{abstract}
Giant diffusion, where the diffusion coefficient of a Brownian particle in a periodic potential with an external force is 
significantly enhanced by the external force, 
is a non-trivial non-equilibrium phenomenon.  We propose a simple stochastic model of giant diffusion, which is based on a biased continuous-time random walk (CTRW) with flight time. {\color{black}By introducing a flight time representing traversal dynamics, we derive the diffusion coefficient using renewal theory and demonstrate its universal peak behavior under various periodic potentials, especially in low-temperature regimes.}
Giant diffusion is universally observed in the sense that there is a peak of the 
diffusion coefficient for any tilted periodic potentials  and the degree of the diffusivity is greatly enhanced 
especially for low-temperature regimes. 
The biased CTRW models with flight times are applied to diffusion under three tilted periodic potentials. 
Furthermore, the temperature dependence of the maximum diffusion coefficient and the external force that attains the maximum are presented for diffusion under
 a tilted sawtooth potential.
\end{abstract}

\begin{document}

\preprint{APS/123-QED}

\title{Universality of giant diffusion  in tilted periodic potentials}

\author{Kento Iida}
\affiliation{%
  Department of Physics, Tokyo University of Science, Noda, Chiba 278-8510, Japan
}%

\author{Andreas Dechant}
\affiliation{%
  Department of Physics No. 1, Graduate School of Science, Kyoto University, Kyoto 606-8502, Japan
}%

\author{Takuma Akimoto}
\email{takuma@rs.tus.ac.jp}
\affiliation{%
  Department of Physics, Tokyo University of Science, Noda, Chiba 278-8510, Japan
}%

\date{\today}


\maketitle

\section{INTRODUCTION}
A tiny particle immersed in an aqueous solution exhibits a random zigzag motion by collisions with surrounding water molecules \cite{brown1828xxvii, einstein1905molekularkinetischen}. This motion is called Brownian motion. There are two ways to describe the irregular motions. 
One is a stochastic dynamic equation of motion, i.e., the Langevin equation \cite{langevin1908theorie}, which describes the trajectory of a Brownian particle. 
The other is a partial differential equation describing the time evolution of the density, i.e., the diffusion equation \cite{fick1855v}. These two equations are equivalent in the sense 
that the two equations can be derived from each other \cite{ risken1996fokker}. The diffusivity of a Brownian particle can be characterized by the mean square displacement (MSD). 
For normal diffusion, the MSD is proportional to time $t$ \cite{einstein1905molekularkinetischen}. The diffusion coefficient $D$, which is a degree of diffusivity, is 
defined by the slope of the MSD in the long-time limit, i.e., $D\sim \langle x(t)^2\rangle/(2t)$ for $t\to\infty$, where $x(t)$ is the position of a one-dimensional Brownian motion at time $t$ 
with $x(0)=0$ and $\langle \cdot \rangle$ represents the ensemble average. Diffusion coefficient $D_{\rm E}$ is determined by the temperature $T$ and the viscous friction coefficient $\eta$ 
of the aqueous solution:
\begin{align}
	D_{\rm E} = \frac{k_{\rm B} T}{\eta} , 
\end{align}
which is known as the Einstein's relation \cite{chandrasekhar1943stochastic}, where  $k_{B}$ is Boltzmann's constant and $\eta$ is viscous friction coefficient.

The continuous-time random walk (CTRW) is a fundamental stochastic model of diffusion as well as anomalous diffusion, where the MSD does not grow linearly with time $t$, i.e., 
$ \langle x(t)^2\rangle \propto t^\alpha$ with $\alpha\ne 1$ \cite{montroll1965random,metzler2000random}. 
The CTRW is a random walk with continuous random waiting times between jumps, where the waiting times are independent and identically distributed (IID) random variables. When the probability of a one-dimensional random 
walker stepping in the right direction is not equal to $1/2$, i.e., asymmetric random walk, 
it is called a biased CTRW. The CTRW has a deep connection with a random walk on a random energy landscape, i.e., 
the trap model \cite{bouchaud1990anomalous, Bardou2002}. In fact, the trap model with a periodic energy landscape 
corresponds to the CTRWs where the waiting-time distribution is identical. 
Such a diffusion is experimentally constructed and the theory of 
the CTRW can be applied to the diffusion in a periodic potential \cite {dechant2019continuous}. When the energy depth at a site is randomly distributed 
according to an exponential distribution, the waiting-time distribution follows a power-law distribution \cite{bouchaud1990anomalous}. 
In the CTRW and the quenched trap model, anomalous diffusion, ergodicity breaking, and non-self averaging of transport coefficients are observed when the mean waiting time diverges. \cite{bouchaud1990anomalous,metzler2000random, He2008, Miyaguchi2011, Miyaguchi2013, Metzler2014, Akimoto2016, Akimoto2018}. 

Several experimental systems are described by Brownian motions in tilted periodic potentials. For example, a rotational motion of $\rm{F_{1}-ATPase}$, which is 
a molecular motor synthesizing ATP exhibits a thermal motion under a periodic potential and a constant torque can be added experimentally \cite{hayashi2015giant}. 
Therefore, the motor's rotational position can be described by the Brownian motion in a tilted periodic potential. Other examples include the diffusion of ions in simple pendulums \cite{risken1996fokker}, superconductors \cite{dieterich1980theoretical}, and Josephson tunneling Junction \cite{anchenko1969josephson}. Ignoring the inertia term in the Langevin equation, i.e., the overdamped Langevin equation, yields the dynamic equation of the Brownian motion.  
The overdamped Langevin equation under a tilted periodic potential is described by
{\color{black}
\begin{align}
\label{eq:OLE}
&\eta \frac{dx}{dt} = -\frac{dU(x)}{dx}+\sqrt{2\eta k_{\rm B} T } \xi(t) ~,
\end{align}
where $U(x) := V(x) -Fx$ and} $V(x)$ is the periodic potential with period $L$, i.e., $V(x+L)=V(x)$, $F$ is a constant external force, and $\xi(t)$ is a white Gaussian noise 
with delta correlation $\langle \xi(t) \xi(t') \rangle=\delta(t-t')$. Similar systems such as diffusion of active Brownian particles in a tilted periodic potential
as well as diffusion under a soft matter potential are also investigated recently \cite{su2023active,lu2023giant}.

Brownian motion in a tilted periodic potential exhibits a giant increment of the diffusion coefficient 
by external force $F$ compared to that without external force, which is called giant diffusion (GD)
\cite{reimann2001giant,*reimann2002diffusion}. For Brownian motion in tilted periodic potentials, the diffusion coefficient $D$ is computed  using the first passage time (FPT)  statistics \cite{reimann2001giant,*reimann2002diffusion}: 
{\color{black} 
    \begin{align}
        D =
        \frac{L^2}{2}
        \frac{\Delta T_2(x_0 \to x_0+L)}
        {[T_1(x_0 \to x_0+L)]^3}~,
    \end{align}
where $T_n(a \to b)$ represents the $n$th moment of FPT, and  the variance of the first-passage time is given by 
    \begin{align}
        &\langle \Delta T_2(x_0 \to x_0 +L)\rangle \nonumber \\ 
        &=
        T_2(x_0 \to x_0 +L) - [T_2(x_0 \to x_0 +L)]^2
    \end{align}
For $F>0$, the FPT moments can be computed using a well-known analytical form \cite{hanggi1990reaction}
    \begin{align}
        T_{n}(a \to b) 
        =
        &\frac{n}{D_E}
        \int_{a}^{b}dx~e^{\beta U(x)}
        \int_{-\infty}^{x}dy~e^{-\beta U(x)} \nonumber\\ 
        &\times T_{n-1}(y\to b) .
    \end{align}
Thus, the diffusion coefficient can be expressed analytically as 
\cite{reimann2001giant,*reimann2002diffusion}:
\begin{align}
\label{eq:RDC}
D 
= 
\frac
{D_{\rm E} L^{2} \int_{0}^{L} I_{+}^{2}(x) I_{-}(x) dx}
{\qty(\int_{0}^{L}I_{+}(x) dx)^{3}},
\end{align}
where  the functions  $I_{\pm}(x)$  are given by
\begin{align}
I_{\pm}(x) 
=
\int_{0}^{L}e^{\beta \qty(\pm V(x) \mp V(x \mp y) -Fy )}dy. 
\end{align}
However, since the diffusion coefficient is expressed as an integral, it is not explicitly represented as a function of the external force  $F$ . This makes it challenging to reveal the universality of GD in a general context.}

This paper aims to provide a new model of diffusion under a tilted periodic potential
 based on biased CTRWs to investigate a universality of GD. From previous studies \cite {dechant2019continuous}, 
it is shown that the Brownian motion in the periodic potential can be mapped to a CTRW. Therefore, we expect that the Brownian motion in the tilted periodic potential can be mapped 
to a biased CTRW. However, the ordinary biased CTRWs cannot exhibit GD. 
We propose a variation of the biased CTRW model as a model to explain GD.
To construct a model of the GD, we introduce a flight time in the biased CTRW, which is the time for a Brownian particle to move from 
a top to one of its two adjacent bottoms of the potential. 

This paper is organized as follows. In Section II, we describe a biased CTRW with flight times, which is a model of the GD. In Section III, we derive the diffusion coefficient in a 
biased CTRW with flight and show the universality of the GD in tilted periodic potential.
The theoretical results are applied to diffusion in a sawtooth periodic potential. Section IV is devoted to the conclusion. 

\section{MODEL}
A biased CTRW is a CTRW with an asymmetric jump probability, which is a model of diffusion with a constant external force. The convection-diffusion equation can be derived 
by the continuous limit in the biased CTRW \cite{Feller1968}. 
When the second moment of the waiting time diverges, 
 the MSD increases as $t^\alpha$ with $\alpha>1$ in a biased CTRW, which is called a field-induced superdiffusion  \cite{Burioni2014,Gradenigo2016,Akimoto2018b,hou2018biased}. 
In what follows, we denote the probabilities that a particle jumps to the right and the left sites by $P^{+}$ and $P^{-}$, respectively, where $P^{+}+P^{-}=1$. 
In the biased CTRW,  the probabilities are not equal, i.e., $P^{+} \neq P^{-}$. When there is a bias in the system, the probabilities depend on the external force $F$. 

\begin{figure}
\centering
\includegraphics[width = 8.5cm]{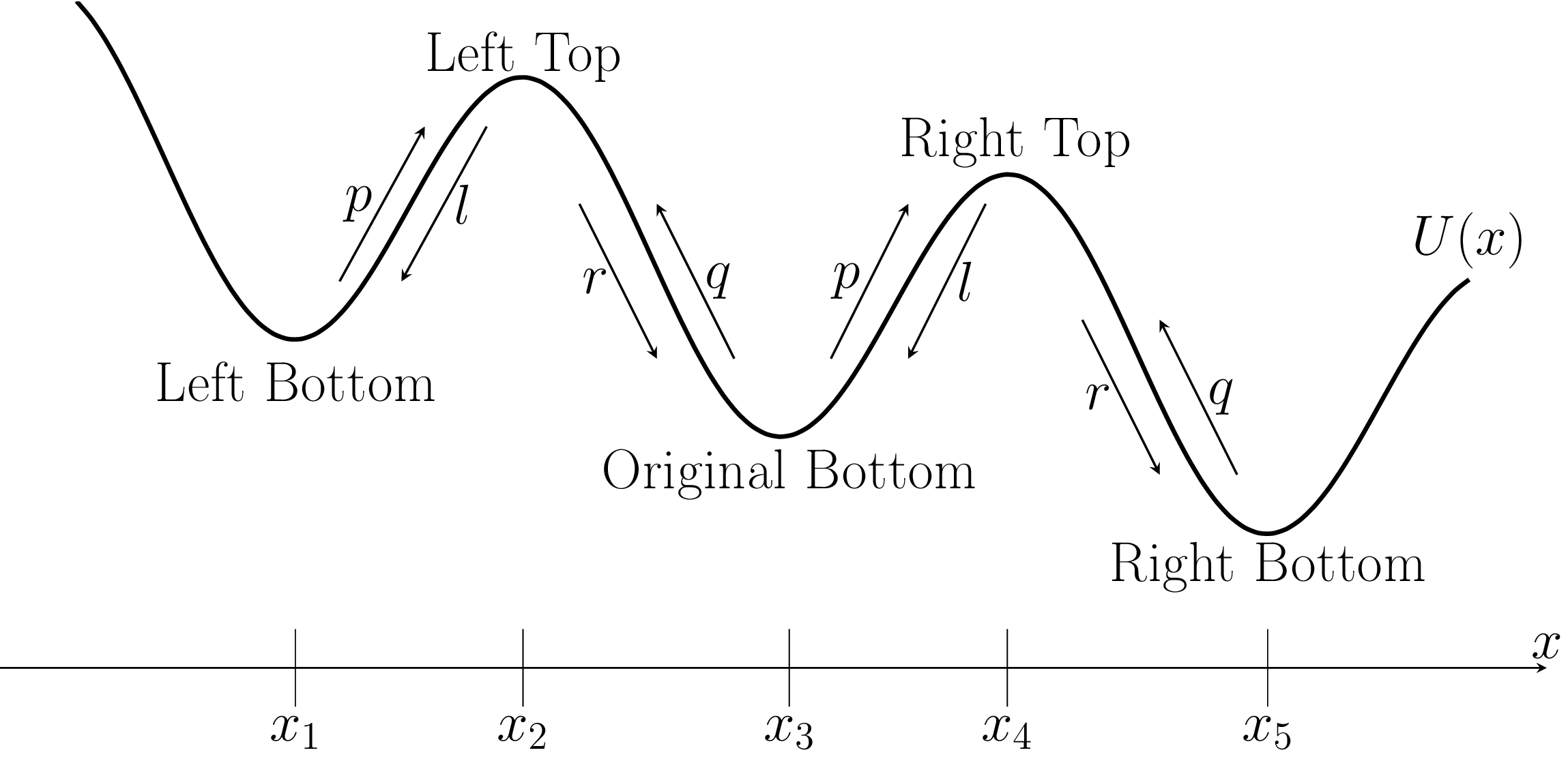}
\caption{Transition probabilities $p,q,r,l$, and the bottom and the top of a tilted periodic potential $U(x)$, where $x_1, x_3, x_5$ are stable points of the potential and 
$x_2$ and $x_4$ are unstable points.
Probabilities $p$ and $q$ are the probabilities of stepping from the bottom to the right and the left top, respectively.  Probabilities $r$ and $l$ are the probabilities of stepping from the top to the right and the left bottom, respectively.}	
\label{fig:periodic}
\end{figure}

We develop a stochastic model of a coarse-grained Brownian motion in a tilted periodic potential $U(x)=V(x)-Fx$, 
where we assume the periodic potential $V(x)$  is symmetric (see Fig.~\ref{fig:periodic}). 
In this coarse-grained model, the particle's position is defined by the position of the nearest neighbor bottom 
 of the potential well. Consequently, the particle's position is discretized, yielding a lattice random walk. 
 A comparable coarse-grained model is also investigated by Lindner et al.~\cite{lindner2001optimal}.
 In what follows, we assume $F>0$ without loss of generality because $V(x)$ is symmetric. In other words, $P_+>P_-$.
 

{\color{black}
In standard biased CTRWs, jumps are typically assumed to occur instantaneously once the waiting time elapses. This assumption holds when the waiting time is significantly longer than the time required for a Brownian particle to traverse from the top to the bottom of the potential. However, this approximation is not valid in general, especially in systems where the traversal time is comparable to the waiting time. To address this limitation, we introduce the flight time  $\theta$, defined as the time it takes for a Brownian particle to move from the top of a potential barrier to one of its neighboring bottoms (see Fig. 2).

The concept of flight time is essential for describing giant diffusion (GD) and represents a key difference from Lindner's model \cite{lindner2001optimal}. In our approach, the total waiting time  $\tau$  is defined as the minimum of two random variables:  $\tau_+$, the time to reach the right peak before the left peak, and $\tau_-$, the time to reach the left peak before the right peak. Both  $\tau$  and  $\theta$  are treated as random variables. 
The net waiting time  $T$, representing the total time required for a Brownian particle to move from one potential bottom to the next, is given by the sum of the waiting time and flight time, i.e.,  $T = \tau + \theta$.
By incorporating the flight time into the CTRW framework, our model provides a more realistic representation of Brownian motion in tilted periodic potentials. This extension is crucial for understanding and predicting the universal behavior of giant diffusion, especially in regimes where the flight time is not negligible.

It is important to note that in symmetric tilted periodic potentials, the PDFs of the waiting times  $\tau_+$  and  $\tau_-$  can be equal under the assumption of independent escape processes, as shown in Lindner's model \cite{lindner2001optimal}. This equality arises because the escape rates to the left and right depend solely on the energy barriers and the external bias, assuming memoryless escape dynamics.
However, in our model, the waiting times  $\tau_+$  and  $\tau_-$  are not assumed to be equal due to the inclusion of flight time and the consideration of a coupled escape process. Specifically, the particle must traverse the potential landscape after reaching the top, introducing additional correlations and memory effects. 
Therefore, while our approach generalizes Lindner's result, it also extends beyond it by incorporating the flight time and treating the waiting-time distributions as distinct random variables.

}

\begin{figure}
\centering
\includegraphics[width = 8.5cm]{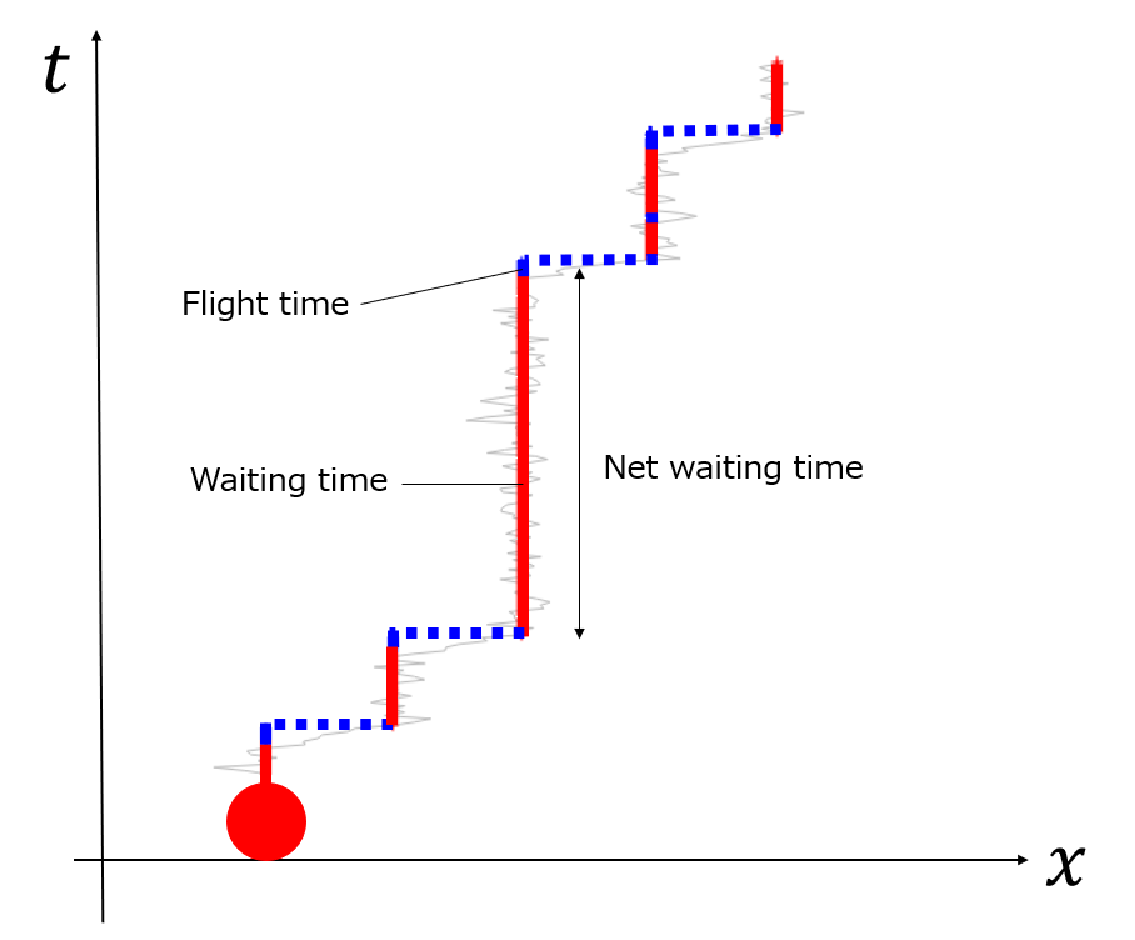}
\caption{Schematic figure of a biased CTRW with flight times. The thin line is a trajectory of a Brownian motion under a tilted periodic potential. 
The solid line is a trajectory of a biased CTRW with flight times, where the net waiting time is a sum of waiting times and flight times.}
\label{fig:BCTRW}
\end{figure}

\section{RESULT}
\subsection{Transition probabilities}

Here, we analytically estimate the transition probabilities of a Brownian particle in a tilted periodic potential (see Fig.~\ref{fig:periodic}). 
Transition probabilities $p$ and $q$ are the probabilities of stepping from the bottom
to the right and the left top, respectively. Transition probabilities $r$ and $s$ are the probabilities of stepping from the top 
to the right and the left bottom, respectively. The transition probabilities $p,q,r, l$ are obtained analytically for a general potential \cite{gardiner2009stochastic}. 
Using stable and unstable points of $U(x)$, we have the transition probabilities $p$ and $q$: 

	\begin{align}
	p 
	= 
	\frac
	{\int_{x_{2}}^{x_{3}} e^{\beta U(y)} dy} 
	{\int_{x_{2}}^{x_{4}} e^{\beta U(y)} dy} 
	,
	q 
	= 
	\frac
	{\int_{x_{3}}^{x_{4}} e^{\beta U(y)} dy} 
	{\int_{x_{2}}^{x_{4}} e^{\beta U(y)} dy} ,
	\label{eq:pq}
	\end{align}
where $\beta$ is the inverse temperature, i.e., $\beta=1/(k_{\rm B}T)$, $x_2$ and $x_4$ are unstable points, $x_3$ is a stable point of potential $U(x)$. 
Moreover, the transition probabilities $r$ and $l$ are also given by 
	\begin{align}
	r 
	= 
	\frac
	{\int_{x_{3}}^{x_{4}} e^{\beta U(y)} dy}
	{\int_{x_{3}}^{x_{5}} e^{\beta U(y)} dy} 
	,
	l
	= 
	\frac
	{\int_{x_{4}}^{x_{5}} e^{\beta U(y)} dy} 
	{\int_{x_{3}}^{x_{5}} e^{\beta U(y)} dy} ,
	\label{eq:rs}
	\end{align}
where $x_5$ is a stable point. 
Because $U(x)$ is a periodic tilted potential, probabilities $r$ and $l$ are also described by 
	\begin{align}
	r 
	= 
	\frac
	{\int_{x_{1}}^{x_{2}} e^{\beta U(y)} dy}
	{\int_{x_{1}}^{x_{3}} e^{\beta U(y)} dy} \ 
	,
	l 
	= 
	\frac
	{\int_{x_{2}}^{x_{3}} e^{\beta U(y)} dy} 
	{\int_{x_{1}}^{x_{3}} e^{\beta U(y)} dy} ,
	\end{align}
where $x_1$ is a stable point. 
From Eqs. \eqref{eq:pq} and \eqref{eq:rs}, equalities $p+q=1$ and $r+l=1$ are satisfied.

\subsection{Diffusion coefficient}



{\color{black} 
In CTRWs, a step from the bottom of the potential to either the left or right neighboring bottom typically occurs immediately after the assigned waiting time elapses. However, in Brownian motion within a tilted periodic potential, reaching a top does not guarantee that the particle will move directly to the next bottom. After reaching a top, the particle can return to the original bottom due to thermal fluctuations. As a result, the effective waiting time includes multiple unsuccessful attempts before a successful jump, making it longer than the first waiting time.

To describe this process, we define the following probability density functions (PDFs):
	$\phi(t)$: The PDF of the time required for a Brownian particle to reach a top and return to the original bottom without reaching the neighboring bottom.
	$\chi(t)$: The PDF of the time required for the particle to move from the original bottom to one of the neighboring bottoms (either left or right) without returning to the original bottom.
	$\omega(t)$: The PDF of the total time required for the particle to move from the original bottom to the next neighboring bottom after reaching either the left or right top, given by the sum of the waiting time  $\tau$  and the flight time  $\theta$.}

 By a simple calculation, the Laplace transforms of  $\phi(t)$ and $\chi(t)$ with respect to $t$  become 
	\begin{align}
		\wt{\phi}(s) &= (pl +  qr ) \ \wt{\omega}(s) , \\
		\wt{\chi}(s) &= (pr +  ql ) \ \wt{\omega}(s) .
	\end{align}
We denote the Laplace transform of function $f(t)$ with respect to $t$ by $\wt{f}(s)$.
Net waiting time $T$ is the time required for a Brownian particle to move from a bottom to one of the next neighboring bottoms.
When a Brownian particle jumps to the next neighboring well after it moves a top and returns to the original well $n-1$ times,  
net waiting time $T$ is a $n$-times sum of $\tau+\theta$. Therefore, 
the PDF of net waiting time $T$, $\psi(t)$, is given by
 	\begin{align}
	\wt{\psi}(s) 
	=\wt{\chi}(s)  \sum_{n=1}^{\infty} \wt{\phi}^{n-1}(s) 
	=\frac{\wt{\chi}(s)} {1-\wt{\phi}(s)} ,
	\end{align}
where we used the fact that the distribution of the sum of random variables can be expressed by the product of the corresponding Laplace transformations. 
Furthermore, using a formula between the moments of a random variable and the Laplace transform of the PDF, i.e., 
	\begin{align}
	\label{eq:LT}
	\left. E[T^{r}] = (-1)^{r} \frac{d^{r} \wt{\psi}(s)} {ds^{r}} \right|_{s=0} ,
	\end{align}
	where $E[\cdot ] $ represents the expectation value, 
we have the mean $\mu$ and the variance $\sigma^{2}$ of the net waiting time: 
	\begin{align}
	\label{eq:mu}
	\mu
	&= \frac{E[\tau]+E[\theta]}{pr+ql} , \\
	\label{eq:sigma} 
	\sigma^{2}
	&= \frac{V[\tau]+V[\theta]} {pr+ql} + \frac{pl+qr}{(pr+ql)^{2}} (E[\tau]+E[\theta])^{2} ,
	\end{align}  
where  $V[\tau]$ and $V[\theta]$ are the variances of  $\tau$ and $\theta$, respectively. Moreover, probability $P^{+}$ of moving from a bottom to the next right bottom
is given by 
	\begin{align}
	\label{eq:pplus}
	P^{+} 
	=pr \sum_{n=1}^{\infty} (pl+qr)^{n-1}  
	=\frac{pr}{pr+ql}  . 
	\end{align}
The probability $P^{+}$ can be also derived from the detailed balance equation.  The detailed balance equation yields
\begin{align}
\frac{\gamma(z \to z')}{\gamma(z' \to z)} = e^{\beta(U(z') - U(z))},
\end{align}
where $\gamma(z \to z')$ is the rate (probability per unit time) of transition from point $z$ to $z'$.
It follows that the ratio of rates $\gamma(x_{0} \to x_{0}+L)$ and $\gamma(x_{0}+L \to x_{0})$  becomes
\begin{align}
\frac{\gamma(x_{0} \to x_{0}+L)}{\gamma(x_{0}+L \to x_{0})} 
=
e^{\beta F L}~.
\end{align}
The rates $\gamma(x_{0} \to x_{0}+L)$ and $\gamma(x_{0}+L \to x_{0})$ can be represented by the mean net waiting time $\langle T\rangle$ and probabilities $P^{+}$ and $P^{-}$: 
\begin{align}
\gamma(x_{0} \to x_{0}+L)  = \frac{P^{+}}{\langle T\rangle},\nonumber\\
\gamma(x_{0}+L \to x_{0}) = \frac{P^{-}}{\langle T\rangle}.
\end{align}
Therefore, we have $P^{-} = P^{+} e^{-\beta F L}$. Substituting this relation into $P^{+}+P^{-}=1$, the external force dependence of $P^{+}$ is given by
\begin{align}
\label{eq:upp}
P^{+} 
= 
\frac
{\exp(\beta F L/2)}
{\exp(\beta F L/2) + \exp(-\beta F L/2)}.
\end{align}

We calculate the diffusion coefficient of a biased CTRW with flight times, where the lattice constant is $L$. We denote the number of jumps of a random walker from a bottom to the next neighboring bottom until time $t$ by $N_{t}$. Net waiting times $T$ are IID random variables. Therefore, $N_t$ is described by a renewal process \cite{cox1967renewal}. 
  By the renewal theory, the  mean $\langle N_{t} \rangle$ and the variance $\langle \Delta N^{2}_{t} \rangle$ of $N_t$ are given by 
	\begin{align}
		\label{eq:N}
		\langle	N_{t} \rangle &= \frac{1}{\mu} t + O(1) , \\
		\label{eq:N2}
		\langle	\Delta N^{2}_{t} \rangle &= \frac{\sigma^{2} }{\mu^{3}} t + O (1) ,
	\end{align}
in the long-time limit. 
The results are valid when $\mu$ and $\sigma^2$ are finite. The variance of the displacement in a biased CTRW with flight times is given by 
	\begin{align}
		\label{eq:varx}
		\langle	\Delta x^{2}_{t} \rangle 
		=  4P^{+}P^{-} L^{2} \langle	N_{t} \rangle 
		+  (P^{+}- P^{-})^{2}  L^{2} \langle	\Delta N_{t}^{2} \rangle .
	\end{align}
	We define the diffusion coefficient $D$ as 
	\begin{align}
		\label{eq:DCD}
		D := 
		\lim_{t \to \infty}	
		\frac{\langle	\Delta x^{2}_{t} \rangle}{2t} .
	\end{align}
Substituting Eqs. \eqref{eq:N}, \eqref{eq:N2}, \eqref{eq:varx} into Eq.~\eqref{eq:DCD} yields
	\begin{align}
		\label{eq:DC}
		D 
		= \frac{L^{2}}{2} 
		\qty(
		\frac{4P^{+}(1-P^{+})}{\mu} 
		+ \frac{(2P^{+}-1)^{2} \sigma^{2}} {\mu^{3}}
		). 
	\end{align}
{\color{black} 
This result is exactly the same as that derived by Lindner et al. \cite{lindner2001optimal}. 
The equivalence arises because the renewal theory approach used here accounts for the key statistical properties of the waiting time and flight time, which also underpin Lindner's analysis. Moreover, the correspondence between Eq.~(\ref{eq:RDC}) and Eq.~(\ref{eq:DC}) has been demonstrated analytically in their work.

However, while the final expression for the diffusion coefficient matches, our approach provides additional physical insight by explicitly incorporating the flight time  $\theta$, which was not considered in Lindner's original model. The inclusion of flight time allows us to capture essential aspects of the dynamics that are otherwise overlooked, such as the traversal time from the top of the potential barrier to the adjacent bottom. This contribution plays a critical role in determining the total net waiting time  $T = \tau + \theta$, where  $\tau$  represents the waiting time to reach the top of the potential barrier.

It is worth noting that, although the transition probabilities  $P_\pm$, the mean waiting time  $\mu$, and the variance $ \sigma^2$  in Eq. (\ref{eq:DC}) must be computed numerically, our expression reveals the underlying stochastic mechanism governing giant diffusion. By treating the waiting and flight times as distinct random variables, our model clarifies how transitions between wells contribute to the overall diffusion process. This insight highlights the physical mechanism driving giant diffusion and explains its universality, especially in the low-temperature regime.}
	
\subsection{Universality of giant diffusion}

{\color{black} We investigate how giant diffusion (GD) emerges in the biased CTRW model with flight times by analyzing the diffusion coefficient's dependence on the external force  $F$. We demonstrate that GD universally occurs in periodic potentials, particularly at low temperatures.


\subsubsection{Diffusion Behavior at Small External Force}

Consider a symmetric periodic potential where the barrier energy $\Delta E$ represents the energy difference between a minimum and the adjacent maximum in the tilted potential  $U(x)$. When  $F$  is small, such that $FL/2 \ll \Delta E_{0}$, where $\Delta E_{0}$ is the barrier energy at  F = 0 , the escape process follows the Arrhenius law:
$\langle \tau \rangle \propto e^{\beta \Delta E}$, where $\quad \Delta E = \Delta E_0 - FL/2$.
The waiting time $\tau$ follows an exponential distribution with variance  $V[\tau] = E [\tau]^2$ for low temperatures, i.e. $\beta \Delta E \gg 1$ \cite{hanggi1990reaction}. 
}
For $\beta \Delta E \gg 1$, $E[\tau]$ is much larger than $E[\theta]$. It follows that the variance of net waiting time is approximately equal to the square 
of the mean net waiting time, i.e., $\sigma^{2} \cong \mu^{2}$. 
Using Eq.~(\ref{eq:DC}) and $\sigma^{2} \cong \mu^{2}$, we have the diffusion coefficient $D_{\rm{small}}$ for small external force:
	\begin{align}
	D_{\rm{small}} \simeq \frac{L^{2}}{2 \mu} . 
	\end{align}
This relation can also be obtained directly from Eq.~(\ref{eq:DC}) by using $P_+ \sim 1/2$ for $F\to 0$. 
Therefore, the dependence of $D_{\rm{small}}$ on external force $F$ is determined by $1/\mu$. 
{\color{black}Since $\mu$ decreases as  $F$  increases due to the reduced barrier height $\Delta E$, the diffusion coefficient  $D_{\text{small}}$  increases with  $F$, as}
	\begin{align}
	\frac{d D_{\rm{small}}} {dF} 
	&= -\frac{L^{2}}{2 \mu^{2}} \frac{d \mu}{d F} > 0 .
	\end{align}
Thus, the diffusion coefficient $D$ increases as $F$ increases when the external force is small.

\begin{figure}
	\centering
	\includegraphics[width = 8.6cm]{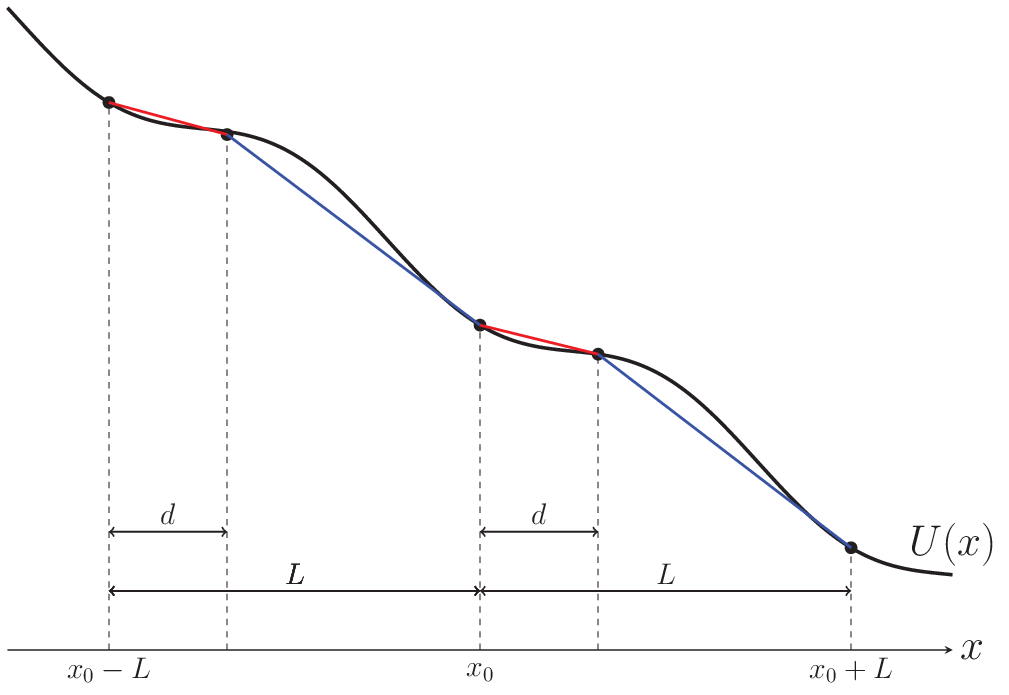}
	\caption{Coarse graining of a tilted periodic potential. The slope of the red straight line is $-a^{-}$, and the slope of the blue straight line is $-a^{+}$. 
	Point $d$ is in between 0 and $L$.}
	\label{fig:2ad}
\end{figure}

\subsubsection{Diffusion Behavior at Large External Force}

{\color{black}When the external force is large,  $FL/2 \gg \Delta E_0$, the periodic potential becomes negligible compared to the force-driven motion, and the particle effectively undergoes free diffusion biased by the external force.}
Because the contribution of the external force to the Brownian dynamics is much larger than that of the periodic potential, it can be regarded as Brownian motion under a constant external force. However, if we simply consider the Brownian motion under a constant external force, we obtain Einstein's relation, i.e., $D=D_{E}$. 
Therefore, we consider a coarse-grained model of the tilted periodic potential by approximating the tilted potential as a piece-wise linear potential (see Fig.\ref{fig:2ad}). Point $d$ in 
Fig.~\ref{fig:2ad} is an arbitrary coordinate and satisfies $0<d<L$. 
The slopes of the piece-wise linear potential  are given by 
 \begin{align}
 	a^{-} &:= -\frac{V(x_{0}+d)-V(x_{0})-Fd}{d}~
 	{\rm for}~0<\tilde{x}<d~,\\
 	a^{+} &:= -\frac{V(x_{0})-V(x_{0}+d)-F(L-d)}{L-d}~
 	{\rm for}~d<\tilde{x}<L~,
 \end{align}
where $\tilde{x} = x-x_{0}-kL$ and $k$ is integer.  
In diffusion under a constant external force $F_0$, i.e., $F=F_0$ and $V(x)=0$ in Eq.~(\ref{eq:OLE}), mean $\mu_{0}$ and variance $\sigma^{2}_{0}$ of the time for a Brownian particle
to move a distance $y$ 
is given by \cite{molini2011first}
	\begin{align}
	\mu_{0} = \frac{\eta y}{F_{0}} ,~  
	\sigma^{2}_{0} = \frac{2 D_{\rm E}  \eta^{3}y}{F_{0}^{3}} . 
	\end{align}
When the external force is large, there is no well in the tilted periodic potential. Moreover, $P_+ \cong 1$ for large $F$. 
In the coarse-grained piece-wise linear potential, the mean and variance of the net waiting time $T$ become
	\begin{align}
	\mu &= \eta \qty(\frac{d}{a^{-}} + \frac{L-d}{a^{+}}), \\  
	\sigma^{2} &= 2D_{E}\eta^{3} \qty( \frac{d}{(a^{-})^{3}} + \frac{L-d}{(a^{+})^{3}}) . 
	\end{align}
It follows that the diffusion coefficient $D_{\rm{large}}$ for large $F$ is given by
	\begin{align}
	D_{\rm{large}} 
	\simeq \frac{L^{2}}{2} \frac{\sigma^{2}}{\mu^{3}}
	= 
	\frac{L^{2}}{\beta \eta} \frac{(a^{+})^{3}d + (a^{-})^{3}(L-d)}
	{\qty(a^{+}d + (L-d)a^{-})^{3}} . 
	\end{align}
The derivative of $D_{\rm{large}}$ with respect to $F$ becomes negative:
	\begin{align}
	\frac{d D_{\rm{large}}}{dF} 
	=
	\frac{3\qty(a^{-} - a^{+})^{2} (a^{-}+a^{+}) d (d-L)}{(da^{+}+(L-d)a^{-})^{4}}
	\frac{4}{\beta \eta} <0 ,
	\end{align}
where we used the fact that $da^{\pm}/dF = 1$. 
When the external force is large, $D$ decreases as $F$ increases and converges to $D_E$ as $F\to \infty$. 
At large forces, the potential wells effectively vanish, reducing the system to free diffusion with a bias. The diffusion coefficient scales inversely with  $F$  due to the piecewise linear traversal dynamics.
Therefore, we find that $D$ increases with small $F$ and decreases with large $F$. 

{\color{black}
\subsubsection{Emergence of a Maximum}

Since the diffusion coefficient  decreases at large  $F$  regardless of the potential's symmetry, there must be a maximum at an intermediate force. This maximum diffusion peak defines giant diffusion and occurs universally even in asymmetric potentials. The universality of giant diffusion extends beyond symmetric potentials to include asymmetric periodic potentials. While the exact functional form of the diffusion coefficient depends on the specific shape of the potential, its qualitative behavior remains the same. 
The emergence of a maximum in the diffusion coefficient is a direct result of the competing effects of decreasing escape times at small  $F$  and force-dominated motion at large  $F$. This non-monotonic behavior guarantees the existence of a peak regardless of the potential's specific shape.
Moreover, the diffusion coefficient at the peak is significantly enhanced especially for low temperature regimes. This is because $D_E$ is much larger than $D$ at $F=0$. 

\subsubsection{ Universality of the Maximum}

The maximum diffusion coefficient is particularly enhanced at low temperatures, where thermal fluctuations are suppressed, and escape dynamics become exponentially sensitive to the energy barriers. This scaling behavior holds regardless of the potential’s shape, ensuring that a maximum must exist for any periodic potential.

At small external forces  $F$, the particle's dynamics are dominated by thermal activation, with escape rates determined by the energy barriers  $\Delta E_{\pm}$. Interestingly, 
for some asymmetric periodic potentials, the escape time may increase as  $F$  increases if the external force tilts the potential, increasing the distance to the next potential barrier or the steepness of the potential. This counterintuitive behavior arises from the interplay between the force-driven tilt and the intrinsic structure of the potential. However, as  $F$  continues to grow, the external force eventually lowers the barrier height in the forward direction, reducing the escape time exponentially. This results in an overall non-monotonic behavior of the escape time with respect to  $F$, driving the diffusion coefficient upward after an initial increase in the escape time.

At large external forces, the dynamics become force-dominated, with the periodic potential playing a negligible role. In this regime, the diffusion coefficient follows the universal scaling: $D_{\text{large}} \propto \frac{1}{F}$.
This force-dependent scaling holds regardless of the potential's shape, meaning that the diffusion coefficient must exhibit a maximum at an intermediate force, defining giant diffusion. Therefore, the universality of giant diffusion extends to both symmetric and asymmetric periodic potentials, with the precise location and height of the maximum depending on the specific energy landscape.
}

\subsection{Giant diffusion in three tilted periodic potentials}

\begin{figure}
	\centering
	\includegraphics[width = 8.6cm]{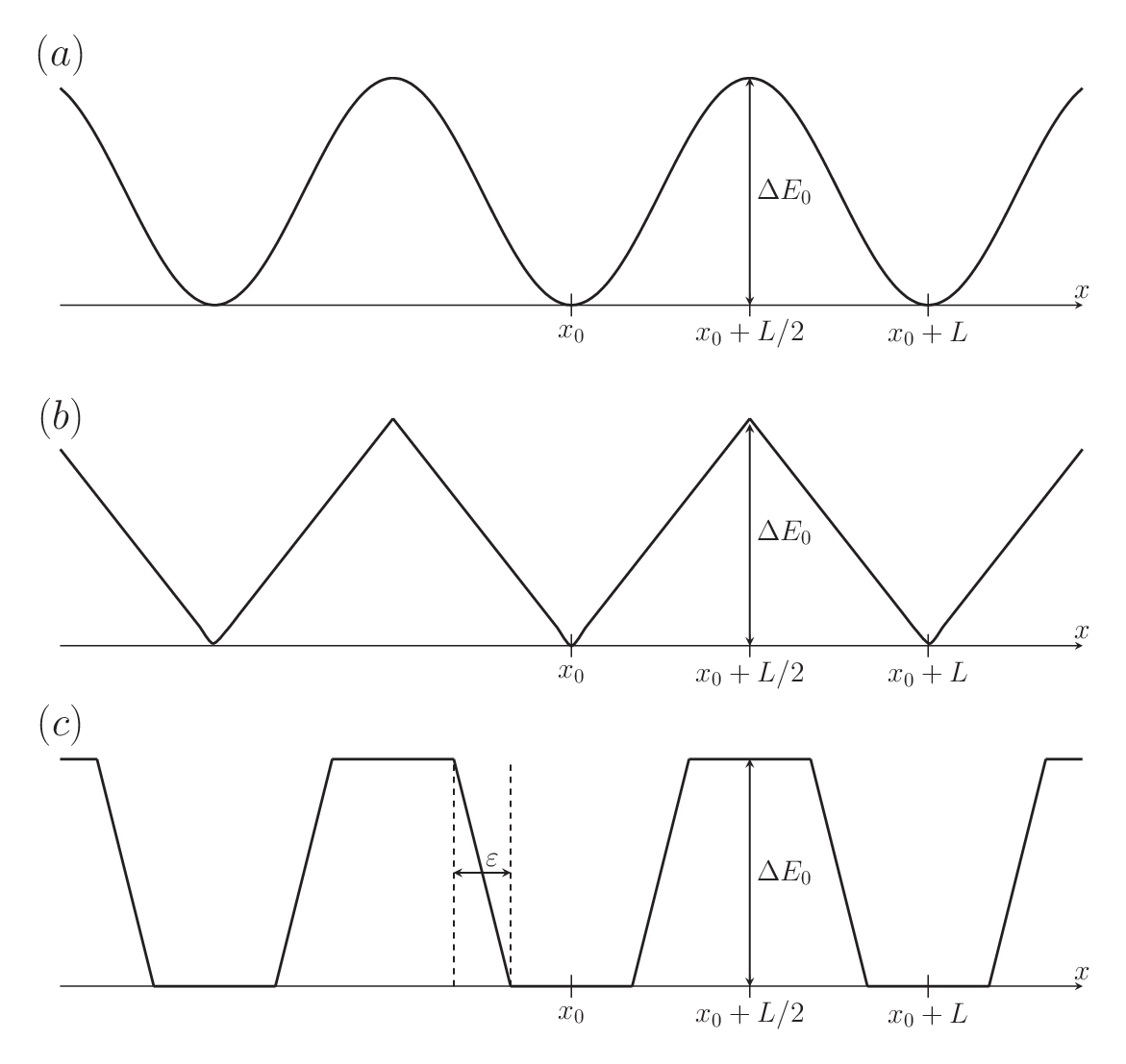}
	\caption
	{Three symmetric periodic potentials, where the difference between the minimum and maximum of the potential is set to be $\Delta E_{0}$. 
	$(a)$~Sine potential.
	$(b)$~Sawtooth potential. 
	$(c)$~Isosceles-trapezoid potential. 
	}
	\label{fig:3models}
	\end{figure}

We show giant diffusion for three symmetric periodic potentials as concrete examples. 
We note that $\Delta E_{0}$ is the barrier energy when there is no external force, and $L$ is the period.
The first example is a sine potential (see Fig.~\ref{fig:3models}a) :
\begin{align}
  		V(x)=-\frac{\Delta E_{0}}{2} \cos{\qty(\frac{2 \pi }{L} x)}.
\end{align}
The second example is a sawtooth potential (see Fig.~\ref{fig:3models}b):
\begin{align}
  		V(x)=
  		\frac{2 \Delta E_{0}}{L} \abs{x-kL} 
  		~{\rm for} ~\abs{x-kL}<L/2 
\end{align}
where $k$ is integer. The sawtooth potential is a piecewise linear approximation of the sine potential. 
In numerical simulations, the force at $x=kL$ is set to be  0.
The third example is an isosceles-trapezoid potential (see Fig.~\ref{fig:3models}c):
\begin{widetext}
\begin{align}
  		V(x)=
  		\begin{cases}
    	0
    	& {\rm for}~\abs{x-kL} < L/4 - \varepsilon/2 ,\\
	\\
    	\dfrac{\Delta E_{0}}{\varepsilon}
    	(\abs{x-kL}-L/4)+\Delta E_{0}/2
    	& {\rm for}~\abs{x-(k\pm1/4)L}<\varepsilon/2,\\
	\\
    	\Delta E_{0}
    	& {\rm for}~\abs{x-kL} > L/4 + \varepsilon/2,
    	\end{cases}
\end{align}
\end{widetext}
where we assume $L>\varepsilon/2$.
The isosceles trapezoid potential has straight lines with finite slopes $\pm \Delta E_{0}/\varepsilon$ in intervals with length $\varepsilon$ (see Fig.~\ref{fig:3models}c). 
In the limit $\varepsilon \to 0$, it becomes a rectangular (square) potential. The isosceles-trapezoid potential is a generalization of the sawtooth potential. 
In particular, when $\varepsilon=L/2$, it becomes a sawtooth potential.

\begin{figure}
	\centering
	\includegraphics[width = 9.0cm]{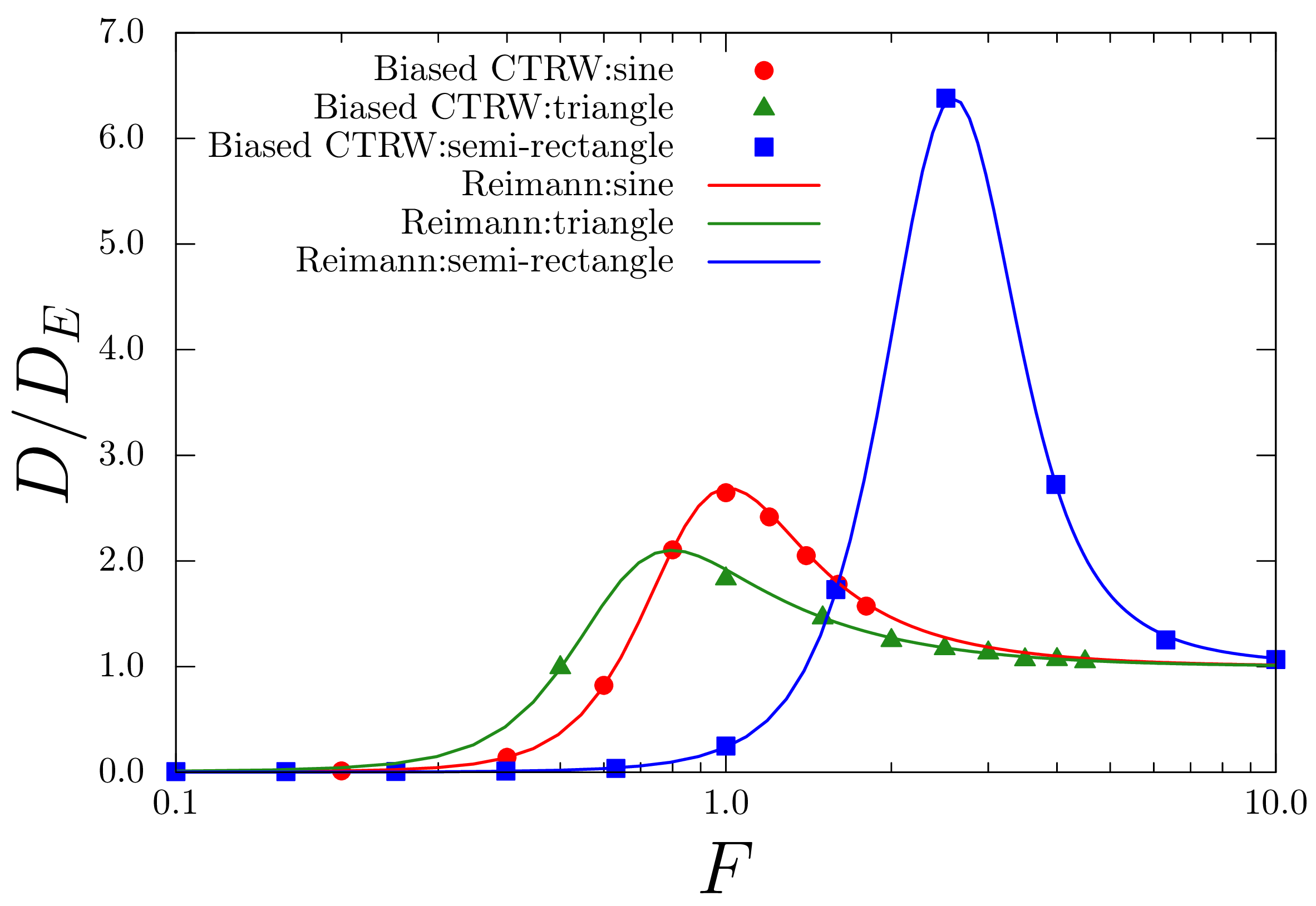}
	\caption{External force dependence of the diffusion coefficient for the three periodic potentials.
	We use $\beta=5, \eta=1, L=2\pi, \Delta E_{0}=2$ for all the potentials, and $\varepsilon=\pi/5$ 
	in the isosceles trapezoidal potential.
	The solid line represents the results of Reimman et al., i.e., Eq.~(\ref{eq:RDC}), by numerical
	integration. The symbols represent the results of a biased CTRW with flight times, i.e., Eq.~(\ref{eq:DC}), where parameters $\mu$ and $\sigma^{2}$ are 
	estimated from simulation of the Langevin equation (\ref{eq:OLE}) and $P^{+}$ is calculated from \eqref{eq:upp}. 
	}
	\label{fig:DC}
	\end{figure}

We apply our theory of the diffusion coefficient for a biased CTRW with flight times to the three periodic potentials by numerical simulations and compare the results with 
  the result of Reimman et al., i.e., Eq.~(\ref{eq:RDC}).  Figure~\ref{fig:DC} shows the diffusion coefficients as a function of external force $F$ for the three periodic potentials. 
  Giant diffusion is observed for the three periodic potentials. Our results of a biased CTRW with flight times are in good agreement with those of Reimman's result. 
  In numerical simulations of the Langevin equation (\ref{eq:OLE}), the simulation was performed by discretizing Eq.~\eqref{eq:OLE}  using Euler's method. 
In addition, white Gaussian noise~$\xi(t)$ was generated using the Box-Muller method. To estimate $\mu$ and $\sigma^2$ in Eq.(\ref{eq:DC}), we calculate 
the first passage time that a trajectory reaches the absorption boundary $x_{0} \pm L$, where $x_{0}$ is the starting point of the trajectory.
For the sine potential, $x_{0}$  is defined by subcritical force $F_{s}=\pi \Delta E_{0}/L$:
\begin{align}
		x_{0}
		=
		\begin{cases}
    	\dfrac{L}{2 \pi} \arcsin \qty(\dfrac{F}{F_{s}})
    	& {\rm for}~0<F<F_{s} \\
	\\
    	\dfrac{L}{4}
    	& {\rm for}~F_{s}<F
    	\end{cases}
    	.
\end{align}
For sawtooth and isosceles-trapezoid potentials, $x_0$ is  $x_{0}=0$. 
	
\subsection{Asymptotic behaviors of the diffusion coefficient}

Here, we show the asymptotic behaviors of the diffusion coefficient in a tilted sawtooth potential for a large- and small-external force and large- and low-temperature limits. 
For the large-$F$ limit, we obtain the Einstein's relation, i.e., $D=D_E$. This is valid for any periodic potentials. 

In a tilted sawtooth potential, expanding  Eq.~\eqref{eq:DC}  near $F=0$ yields
	\begin{align}
	\label{eq:DFS}
	D = \frac{e^{\beta V} V^{2} \beta}{(1-e^{\beta V})\eta} + O(F^{2}) .
	\end{align}
For small-$F$ limit, the diffusion coefficient behaves as $D - D_0 \propto F^2$, where $D_0$ is the diffusion coefficient for $F=0$.  
Because periodic potential $V(x)$ is symmetric, the order of the sub-leading term is $O(F^2)$ in general.  
Furthermore, expanding Eq.~\eqref{eq:DFS} around $\beta=0$ gives
	\begin{align}
	D = \frac{1}{\beta \eta} + O(\beta) + O(F^{2}).
	\end{align}
Therefore, the diffusion coefficient converges to $D_E$, i.e., the Einstein relation holds, in the high-temperature limit even when the external force is small. 
This means that the situation is almost the same as that of free diffusion because large thermal noises overcome the potential. Furthermore, 
for a low-temperature limit $\beta \rightarrow \infty$ in Eq.~\eqref{eq:DFS}, the diffusion coefficient becomes 
	\begin{align}
	D \simeq \frac{V^{2} \beta e^{-\beta V}}{\eta} \rightarrow 0 .
	\end{align}
This means that diffusion rarely occurs at a low-temperature limit.

\begin{figure}
\centering
\includegraphics[width = 8.6cm]{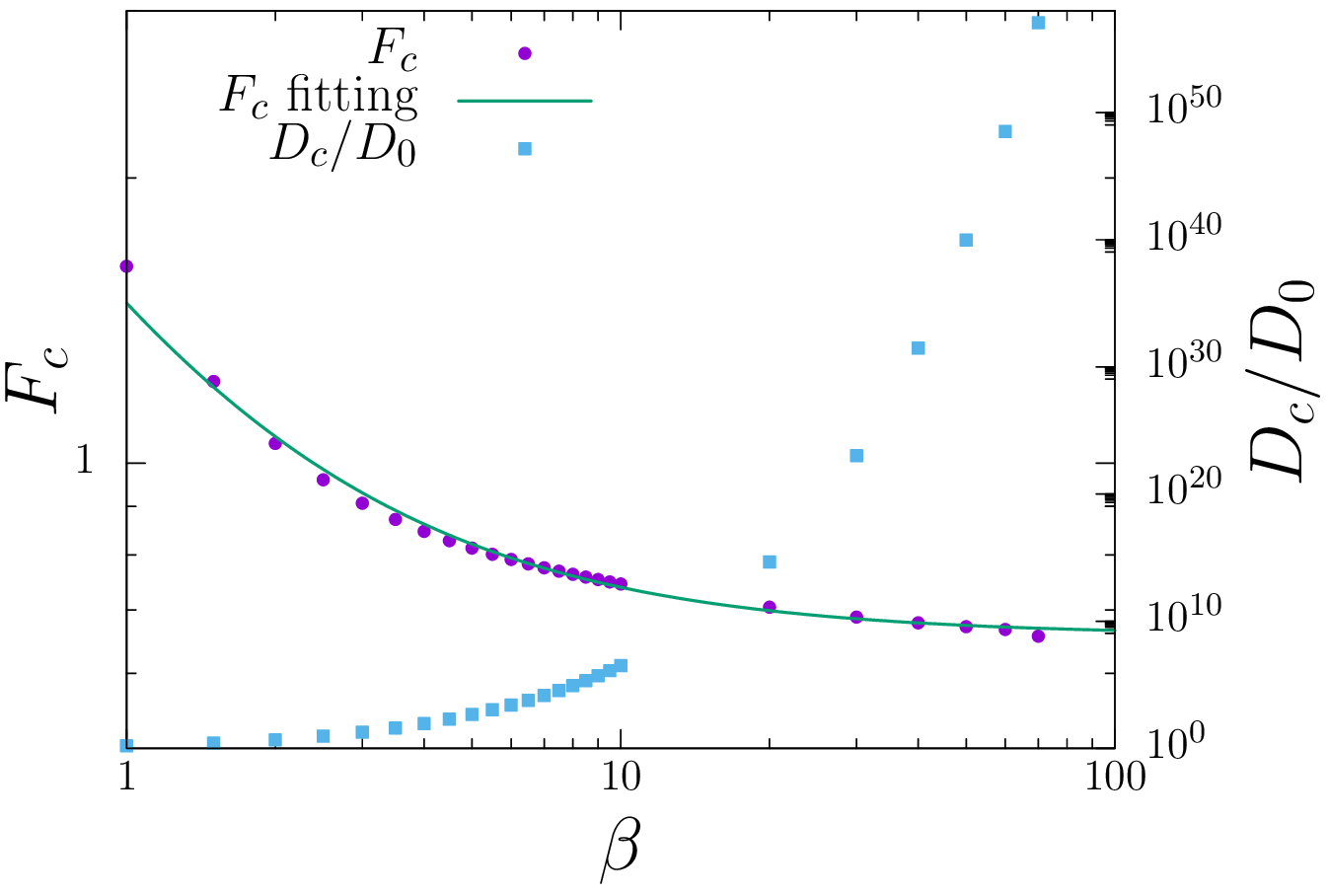}
\caption{Critical force $F_{c}$ and the maximum value diffusion coefficient divided by $D_0$, i.e., $D_c/D_{0}$, as a function of inverse temperature $\beta$ in a biased CTRW with 
flight times, where parameters $\mu, \sigma^2$ and $P_+$ in Eq.~(\ref{eq:DC}) 
are obtained by Eqs.~\eqref{eq:mu}, \eqref{eq:sigma}, \eqref{eq:upp}, \eqref{eq:p_cal}, \eqref{eq:r_cal}, \eqref{eq:Tau1}, \eqref{eq:Tau2}, \eqref{eq:The1}, and \eqref{eq:The2} 
in the sawtooth potential with $L=2\pi$ and $\Delta E_0=2$.  
Circles and squares are the results on numerical simulations for $F_{c}$ and  $D_{c}/D_{0}$, respectively. 
The solid curve is a fitting of critical force  $F_{c}$ by $F_c=A/\beta + B$ with $A=0.817697$ and $B=0.657699$.}
\label{fig:BFC}
\end{figure}

Figure \ref{fig:BFC} shows critical force $F_{c}$ and the maximal diffusion coefficient $D_{c}$ in a tilted periodic sawtooth potential, where the critical force is defined by 
the force at which the diffusion coefficient is maximized. In numerical simulations, we use a biased CTRW with flight times, where parameters $\mu$ and $\sigma^2$ are 
calculated using Eqs.~\eqref{eq:mu}, \eqref{eq:sigma}, \eqref{eq:upp},  \eqref{eq:p_cal}, \eqref{eq:r_cal}, \eqref{eq:Tau1}, \eqref{eq:Tau2}, \eqref{eq:The1}, and \eqref{eq:The2} 
in Appendix~A. We have confirmed that these analytical values are consistent with 
those in numerical simulations.
Critical force $F_{c}$ converges to a constant value as $\beta$ increases. The value of $D_{c}/D_{0}$ significantly increases as $\beta$ increases.

{\color{black}\section{Discussion}
We discuss the significance of introduction of the flight time. 
The flight time  $\theta$ is defined as the time required for a Brownian particle to move from a potential maximum to an adjacent minimum, plays a crucial role in understanding GD. 
In Lindner's model \cite{lindner2001optimal}, the waiting time implicitly includes the traversal time from one potential minimum to the next, effectively assuming instantaneous transitions after escape due to the Markovian approximation. In contrast, our model explicitly separates the waiting time  $\tau$, representing the time to reach the barrier top, from the traversal time  $\theta$, corresponding to the particle's flight across the barrier. This distinction allows us to capture non-Markovian dynamics and account for force-driven traversal effects, which are essential for explaining the universality of giant diffusion in tilted periodic potentials.

By explicitly introducing the flight time  $\theta$, we account for the finite traversal dynamics during the transition from one minimum to the next. The total net waiting time becomes a sum of $\theta$ and  $\tau$.  This correction is essential because the traversal time can become comparable to or even exceed the waiting time in tilted periodic potentials, especially for large external forces.

Additionally, the inclusion of flight time breaks the Markovian assumption, which assumes independent, memoryless escape events. The coupling between the waiting time and the traversal process creates a non-Markovian escape process, generating non-trivial correlations in the particle's dynamics. As a result, the flight time directly influences the mean escape time and its variance, both of which contribute to the diffusion coefficient.

Most importantly, flight time also affects the universality of GD. As the external force increases, the flight time becomes increasingly deterministic due to force-driven traversal, while the waiting time becomes exponentially suppressed. This dynamic interplay creates a non-monotonic dependence of the diffusion coefficient on the external force, ultimately leading to the emergence of a diffusion maximum, even in asymmetric potentials. This insight demonstrates that the flight time mechanism is critical for capturing the universal features of giant diffusion.

}

\section{Conclusion}
We construct a coarse-grained stochastic model of a Brownian motion in a tilted periodic potential by a biased CTRW with flight times.  
We derive the diffusion coefficient of the biased CTRW model with flight times by the renewal theory.  
Compared to the Reimann's result, i.e., Eq.~(\ref{eq:RDC}), our expression of the diffusion coefficient is much more simple. 
Thus, we evaluate an explicit dependence of the diffusion coefficient on the external force for small and large forces. As a result,
we show the universality of giant diffusion in tilted  periodic potentials in the sense that there is at least one peak 
in the diffusion coefficient and the peak value is significantly larger than that without external force. 
We apply our theory to three periodic tilted potentials. We can derive the diffusion coefficients as a function of $F$ if we evaluate 
$\mu$ and $\sigma^2$. 
Our theory can also be applied to asymmetric periodic potentials. 
Even for asymmetric periodic potential, the GD occurs especially for low temperatures. 
Moreover, even in higher dimensional systems, giant 
diffusion can be observed when diffusion directed to the applied force is independent of other directions.
By numerical simulations,
we confirm the validity of our theory and 
the giant diffusion for three tilted periodic potentials. 

\section*{Acknowledgements}
 T.A. was supported by JSPS Grant-in-Aid for Scientific Research (No. C 21K033920). We thank I. Sakai for supporting 
 the research.

\appendix
\section{External force dependence of physical quantities in sawtooth potential}
\begin{figure}[h]
\centering
\includegraphics[width = 9.7cm]{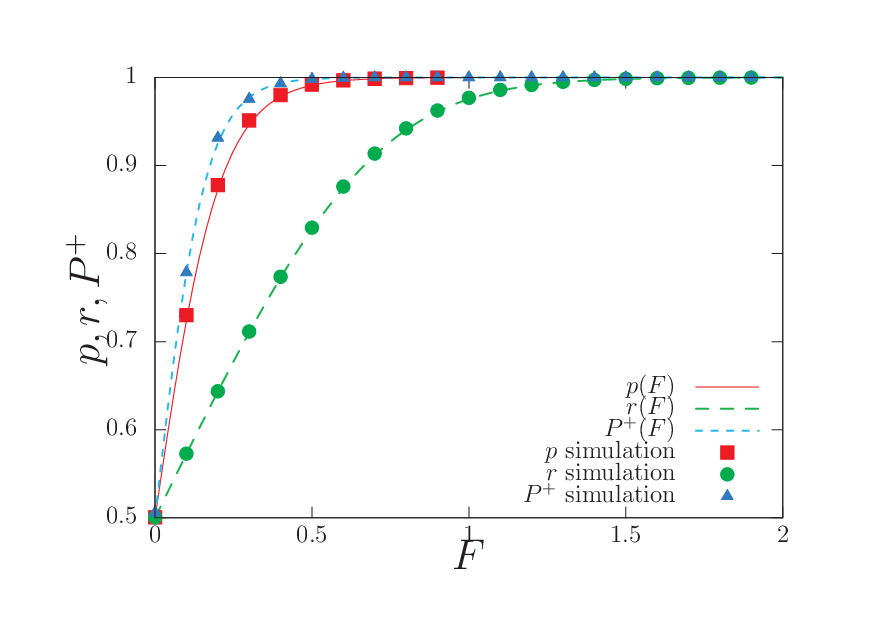}
\caption{External Force dependence of probabilities $p,r,P^{+}$. Symbols are the results of numerical simulations, where we set $\beta=2$. 
The red solid, the green dashed, and the blue dotted lines represent $p(F)$,  $r(F)$, and  $P^{+}$, respectively. }
\label{fig:p0}
\end{figure}

Here, we provide external force dependence of some statistical quantities in a tilted periodic sawtooth potential. In what follows, we use the following notation:
\begin{align}
b^{\pm} := 2V_{0} \pm FL~.
\end{align}
Therefore,  the tilted sawtooth potential $U_{{\rm saw}}(x)$ can be represented as
	\begin{align}
  		U_{{\rm saw}}(x)=
  		\begin{cases}
    	-\frac{b^{+}}{L}x + 2\Delta E_{0}k  
    	&{\rm for}~ (k-1)L/2 \leq x < kL\\
    	+\frac{b^{-}}{L}x - 2\Delta E_{0}k  
    	&{\rm for}~ kL \leq x < (k+1)L/2\\
    	\end{cases}
    	,
	\end{align}
where $k$ is integer.
According to the formulas \eqref{eq:pq}, \eqref{eq:rs}, we obtain the external force dependence of the probabilities $p,r,P^{+}$:
	\begin{align}
	\label{eq:p_cal}
	p 
	&= 
	\frac
	{\frac{1}{b^{+}} (e^{\beta \frac{b^{+}}{2}}-1)}
	{\frac{1}{b^{+}} (e^{\beta \frac{b^{+}}{2}}-1)
	+\frac{1}{b_{-}} (e^{\beta \frac{b^{-}}{2}}-1)} , \\
	\label{eq:r_cal}
	r 
	&= 
	\frac
	{\frac{1}{b^{-}} (e^{\beta \frac{b^{-}}{2}}-1)}
	{\frac{1}{b^{-}} (e^{\beta \frac{b^{-}}{2}}-1)
	+\frac{1}{b_{+}} (1-e^{-\beta \frac{b^{+}}{2}}) e^{\beta \frac{b^{-}}{2}}}~,\\
	\label{eq:Pp}
	P^{+}
	&=
	\frac
	{\exp(\beta F L/2)} 
	{\exp(\beta F L/2 ) + \exp(-\beta F L/2)} . 
	\end{align}
These results agree with numerical results (see Fig.~\ref{fig:p0}). 

\begin{widetext}

The $n$th moment of FPT, which is the time for a Brownian particle to move from the initial position $x=x_{0}$ to the absorbing boundary $x=A$ or $x=B$ (where $A \leq x \leq B$), 
is given by the following formula \cite{gardiner2009stochastic}:
\begin{align}
\label{eq:FPT}
T_{n}(x)
= 
\beta \eta n
\qty(
\frac
{(\int_{A}^{x} e^{\beta U(y)} dy) \int_{x}^{B} e^{\beta U(y)} dy \int_{A}^{y} e^{-\beta U(z)} T_{n-1}(z)dz} 
{\int_{A}^{B} e^{\beta U(y)} dy}
-
\frac
{(\int_{x}^{B} e^{\beta U(y)} dy) \int_{A}^{x} e^{\beta U(y)} dy \int_{A}^{y} e^{-\beta U(z)} T_{n-1}(z)dz}
{\int_{A}^{B} e^{\beta U(y)} dy}) .  
\end{align}	
Therefore, the 1st moment of $\tau$ is represented by 
\begin{align}
\label{eq:Tau1}
E[\tau]
=&
\frac{L^{2} \eta}{\beta} \frac{b^{+} b^{-}}{b^{-}\qty(e^{\beta \frac{b^{+}}{2}}-1) + b^{+} \qty(e^{\beta \frac{b^{-}}{2}}-1)}  \nonumber \\
\times&
\left \{ 
\frac{1} { (b^{+})^{2} b^{-} } \qty(e^{\beta \frac{b^{-}}{2}}-1)  \qty(e^{\beta \frac{b^{+}}{2}} -1  - \beta \frac{b^{+}}{2})  \right.  
\left.  + 
\frac{1} { b^{+} (b^{-})^{2} } \qty(e^{\beta \frac{b^{+}}{2}}-1)  \qty(e^{\beta \frac{b^{-}}{2}} -1  - \beta \frac{b^{-}}{2})
\right \} , 
\end{align}
and the 2nd moment of $\tau$ is given by 
\begin{align}
\label{eq:Tau2}
E[\tau^{2}]
=
2 (p I_{p} - q I_{m}) ,  
\end{align}
where
\begin{align}
\label{eq:Ip}
I_{p}
=&
\frac{L^{4} \eta^{2}}{\beta^{4}} \frac{b^{+} b^{-}}{b^{-}\qty(e^{ \frac{\beta b^{+}}{2}}-1) + b^{+} \qty(e^{ \frac{\beta b^{-}}{2}}-1)}  \nonumber\\
\times&
\left \{
	\frac{1}{(b^{-})^{5}} 
	\qty[
	\qty(e^{\frac{\beta b^{-}}{2}} -1 + \frac{\beta b^{-}}{2}) \qty(e^{\frac{\beta b^{-}}{2}} -1 -\frac{\beta b^{-}}{2}) 
	+\frac{\beta b^{-}}{2} \qty(e^{\frac{\beta b^{-}}{2}} -1)
	+\frac{(\beta b^{-})^{2}}{8} \qty(1-3e^{\frac{\beta b^{-}}{2}}) 
	] 
	\right.  \nonumber \\
	&\left. +
	\frac{1}{(b^{-})^{4} b^{+}}
	\qty(e^{ \frac{\beta b^{+}}{2}}-1)
	\qty[
	\qty(e^{\frac{ \beta b^{-}}{2}} +1 -\frac{\beta b^{-}}{2}) \qty(e^{\frac{\beta b^{-}}{2}}-1-\frac{\beta b^{-}}{2})
	-\qty(1-e^{ \frac{ \beta b^{-}}{2} }) 
	- \frac{\beta b^{-}}{2} e^{ \frac{\beta b^{-}}{2}}
	- \frac{(\beta b^{-})^{2}}{8} 
	]
	\right.  \nonumber\\
	&\left. +
	\frac{1}{(b^{-})^{3} (b^{+})^{2}}
	\qty(e^{ \frac{ \beta b^{+}}{2}} -1  -\frac{\beta b^{+}}{2})
	\qty(2e^{\beta b^{-}} -2e^{\beta \frac{b^{-}}{2}} -\frac{3}{2}\beta b^{-} e^{\frac{\beta b^{-}}{2} } + \frac{\beta b^{-}}{2})
	\right.  \nonumber\\
	&\left. +
	\frac{1}{(b^{-})^{2} (b^{+})^{3}}
	\qty(e^{\frac{\beta b^{+}}{2}} -e^{- \frac{\beta b^{+}}{2} } -\beta b^{+})
	\qty(e^{\frac{\beta b^{-}}{2}}-1)^{2}
	\right.  \nonumber\\
	&\left. +
	\frac{1}{b^{-} (b^{+})^{4}}
	\qty(e^{\frac{\beta b^{+}}{2}} -e^{- \frac{\beta b^{+}}{2} } -2 -\frac{(\beta b^{+})^{2}}{4} )
	\qty(e^{\frac{\beta b^{-}}{2} }-1)
\right \} 
\end{align}
and
\begin{align}
\label{eq:Im}
I_{m}
=&
\frac{L^{4} \eta}{\beta^{3}} \frac{b^{+} b^{-}}{b^{-}\qty(e^{ \frac{\beta b^{+}}{2}}-1) + b^{+} \qty(e^{\frac{\beta b^{-}}{2}}-1)}  \nonumber\\
\times&
\left \{
	\frac{1}{(b^{+})^{5}} 
	\qty[
	\qty(2-e^{ \frac{\beta b^{+}}{2}} -e^{-\frac{\beta b^{+}}{2}}) 
	+\frac{\beta  b^{+}}{2} \qty(1-e^{\frac{\beta b^{+}}{2}})
	+\frac{(\beta b^{+})^{2}} {8}(3+e^{\frac{\beta b^{+}}{2}}) 
	] 
	\right.  \nonumber\\
	&\left. +
	\frac{1}{(b^{+})^{4} b^{-}}
	\qty(e^{ \frac{\beta b^{-}}{2}}-1)
	\qty[
	\qty(1+e^{-\frac{\beta b^{+}}{2}} -2e^{\frac{\beta b^{+}}{2} }) 
	+ \beta b^{+} \qty(1+\frac{1}{2}e^{ \frac{\beta b^{+}}{2}}) 
	- \frac{(\beta b^{+})^{2}} {8}
	]
	\right.  \nonumber\\
	&\left. +
	\frac{1}{(b^{+})^{3} (b^{-})^{2}}
	\qty(e^{ \frac{\beta b^{-}}{2}} -1  -\frac{\beta b^{-}}{2})
	\qty[
	2 \qty(1-e^{ \frac{\beta b^{+}}{2} })
	+ \frac{\beta b^{+}}{2} \qty(1+e^{ \frac{\beta b^{+}}{2}})
	]
\right \} .
\end{align}
In the same way of the calculation of $\tau$, the 1st moment of $\theta$ is given by 
\begin{align}
\label{eq:The1}
E[\theta]
=&
\frac{L^{2} \eta}{\beta} 
\frac{b^{+} b^{-}}{b^{+}\qty(e^{\beta \frac{b^{-}}{2}}-1) + b^{-} e^{2\beta V_{0}} \qty(e^{-\beta \frac{b^{+}}{2}} -e^{-\beta b^{+}})}  \nonumber\\
\times&
\left \{ 
\frac{1}{  b^{+} (b^{-})^{2} } 
e^{2\beta V_{0}} \qty(e^{-\beta \frac{b^{+}}{2}}-e^{-\beta b^{+}})  \qty(e^{-\beta \frac{b^{-}}{2}} -1  - \beta \frac{b^{-}}{2})  \right.
\left.  + 
\frac{1} { (b^{+})^{2} b^{-} } \qty(e^{\beta \frac{b^{-}}{2}}-1)  \qty(e^{-\beta \frac{b^{+}}{2}} -1 + \beta \frac{b^{+}}{2})
\right \} ,
\end{align}
and the 2nd moment of $\theta$ is given by 
\begin{align}
\label{eq:The2}
E[\theta^{2}]
=
2(r J_{p} - s J_{m}) ,
\end{align}
where,
\begin{align}
\label{eq:Jp}
J_{p}
=&
\frac{L^{4} \eta^{2} }{\beta^{2}} 
\frac{b^{+} b^{-}}{b^{+}\qty(e^{\beta \frac{b^{-}}{2}}-1) + b^{-} e^{2\beta V_{0}} \qty(e^{-\beta \frac{b^{+}}{2}} -e^{-\beta b^{+}})}  \nonumber\\
\times&
\left \{
	\frac{e^{2\beta V_{0}}}{(b^{+})^{5}} 
	\qty[
	-\qty(e^{\frac{3 \beta b^{+}}{2}} -2e^{-\beta b^{+}} + \frac{\beta b^{+}}{2})  
	+\frac{\beta b^{+}}{2} \qty(e^{-\beta b^{+}} - e^{\frac{-\beta b^{+}}{2}})
	+\frac{(\beta b^{+})^{2}}{8} \qty(3e^{-\beta b^{+} } - e^{-\frac{\beta b^{+}}{2}}) 
	] 
	\right.  \nonumber\\
	&\left. +
	\frac{1}{(b^{+})^{4} b^{-}}
	\qty(e^{ \frac{\beta b^{-}}{2}}-1)
	\qty[
	\qty(e^{-\beta b^{+}} + e^{-\frac{\beta b^{+}}{2}} -2) 
	+ \frac{3 \beta b^{+}}{2} e^{-\frac{\beta b^{+}}{2}}
	+ \frac{(\beta b^{+})^{2}}{8} 
	]
	\right.  \nonumber\\
	&\left. +
	\frac{e^{2 \beta V_{0}}} {(b^{+})^{3} (b^{-})^{2}}
	\qty(e^{-\frac{ \beta b^{-}}{2}} -1  +\frac{\beta b^{-}}{2})
	\qty[
	2\qty(e^{-\beta b^{+}} -e^{-\beta \frac{3 b^{+}}{2}})
	+\frac{\beta b^{+}}{2} \qty(e^{-\frac{\beta b^{+}}{2} } -3 e^{-\beta b^{+}})
	]
	\right.  \nonumber\\
	&\left. +
	\frac{e^{4\beta V_{0}} }{(b^{+})^{2} (b^{-})^{3}}
	\qty(1 - e^{-\beta b^{-}} -\beta b^{-} e^{-\beta \frac{b^{-} }{2} })
	\qty(e^{-\frac{\beta b^{+}}{2}}-e^{-\beta b^{+}})^{2}
	\right.  \nonumber\\
	&\left. +
	\frac{1}{b^{+} (b^{-})^{4}}
	\qty(e^{\frac{\beta b^{-}}{2}} +e^{- \frac{\beta b^{-}}{2} } -2 -\frac{(\beta b^{-})^{2}}{4} )
	\qty(e^{-\frac{\beta b^{+}}{2} }-e^{-\beta b^{+}})
\right \} ,\\
\label{eq:Jm}
J_{m}
=&
\frac{L^{4} \eta^{2} }{\beta^{2}} 
\frac{b^{+} b^{-}}{b^{+}\qty(e^{\beta \frac{b^{-}}{2}}-1) + b^{-} e^{2\beta V_{0}} \qty(e^{-\beta \frac{b^{+}}{2}} -e^{-\beta b^{+}})} \nonumber\\
\times&
\left \{
	\frac{1}{(b^{-})^{5}} 
	\qty[
	\qty(e^{\beta b^{-}} -2e^{\beta \frac{b^{-}}{2}} +1 ) 
	+\frac{\beta  b^{-}}{2} \qty(e^{\frac{\beta b^{-}}{2}}-1)
	+\frac{(\beta b^{-})^{2}}{8} (1+3e^{\frac{\beta b^{-}}{2}}) 
	] 
	\right.  \nonumber\\
	&\left. +
	\frac{e^{2 \beta V_{0}}}{(b^{-})^{4} b^{+}}
	\qty(e^{-\frac{\beta b^{+}}{2}}-e^{-\beta b^{+}})
	\qty[
	\qty(e^{\frac{\beta b^{-}}{2}} -2e^{-\frac{\beta b^{-}}{2} } +1) 
	- \frac{\beta b^{-}} {2} \qty(2 + e^{-\frac{\beta b^{-}}{2}}) 
	- \frac{(\beta b^{-})^{2}} {8}
	]
	\right.  \nonumber\\
	&\left. +
	\frac{1}{(b^{-})^{3} (b^{+})^{2}}
	\qty(e^{-\frac{\beta b^{+}}{2}} -1  +\frac{\beta b^{+}}{2})
	\qty[
	2
	- 2 e^{ \frac{\beta b^{-}}{2} }
	+ \frac{\beta b^{-}}{2} \qty(1+e^{\frac{\beta b^{-}}{2}})
	]
\right \} .
\end{align}

\end{widetext}




%

\end{document}